\documentclass[prl,letterpaper,superscriptaddress,twocolumn,10pt,aps,notitlepage,nofootinbib,nobibnotes]{revtex4-1}

\usepackage{amsfonts}
\usepackage{amsmath}
\usepackage{graphicx}
\usepackage{hyperref}
\usepackage{anyfontsize}
 \usepackage{ booktabs}
\usepackage{ulem}
\usepackage{lineno}
\usepackage[dvipsnames]{xcolor}
\usepackage{cancel}
\usepackage[utf8]{inputenc}
\usepackage{epsfig,graphics,rotate,color}
\usepackage[dvipsnames]{xcolor}

\hypersetup{colorlinks, linkcolor = [rgb]{0,0.0,0.5}, citecolor = [rgb]{0,0.0,0.5}, urlcolor = [rgb]{0,0.0,0.5}}

\def\be{\begin{equation}}
\def\ee{\end{equation}}
\def\bea{\begin{eqnarray}}
\def\eea{\end{eqnarray}}

\def\be{\begin{equation}}
\def\ee{\end{equation}}
\def\bea{\begin{eqnarray}}
\def\eea{\end{eqnarray}}
\def\bec{\begin{center}}
\def\enc{\end{center}}

\begin{document}

\preprint{SLAC-PUB-17500}

\title{\Large Diffractive Dissociation of Alpha Particles as a Test of Isophobic Short-Range Correlations inside Nuclei}

\newcommand*{\SLAC}{SLAC National Accelerator Laboratory, Stanford University, Stanford, CA 94309, USA}\affiliation{\SLAC}
\newcommand*{\COSTARICA}{Laboratorio de F\'isica Te\'orica y Computacional, Universidad de Costa Rica, 11501 San Jos\'e, Costa Rica}\affiliation{\COSTARICA}
\newcommand*{\CHILE}{Departamento de  F\'isica y Centro Cient\'ifico Tecnol\'ogico de Valpar\'aiso-CCTVal, Universidad T\'ecnica Federico Santa Mar\'ia, Casilla 110-V, Valpara\'iso, Chile}\affiliation{\CHILE}

\author{Jennifer Rittenhouse West}\email{jrwest@slac.stanford.edu}\affiliation{\SLAC}
\author{Stanley~J.~Brodsky}\email{sjbth@slac.stanford.edu}\affiliation{\SLAC} 
\author{Guy~F.~de~T\'eramond}\email{gdt@asterix.crnet.cr}\affiliation{\COSTARICA}
\author{Iv\'an~Schmidt}\email{ischmidt@fis.utfsm.cl}\affiliation{\CHILE}  


\begin{abstract}

The CLAS collaboration at Jefferson Laboratory has compared nuclear parton distributions for a range of nuclear targets and found that the EMC effect measured in deep inelastic lepton-nucleus scattering has a strongly ``isophobic" nature.  
This surprising observation suggests short-range correlations between neighboring $n$ and $p$ nucleons in nuclear wavefunctions that are much stronger compared to $p-p$ or $n-n$ correlations.  In this paper we propose a definitive experimental test of the nucleon-nucleon explanation of the isophobic nature of the EMC effect: the diffractive dissociation on a nuclear target $A$ of high energy $\rm ^4He$ nuclei to pairs of nucleons $n$ and $p$ with high relative transverse momentum, $\alpha + A \to  n + p  +  A'  + X $.  The comparison of $n-p$ events with $p-p$ and $n-n$ events directly tests the postulated breaking of isospin symmetry.  The experiment also tests alternative QCD-level explanations for the isophobic EMC effect.  In particular it will test a proposal for hidden-color degrees of freedom in nuclear wavefunctions based on isospin-zero $[ud]$ diquarks.
\end{abstract}


\maketitle

Diffractive dissociation of relativistic hadrons is an important tool for probing hadron structure. The fundamental quark-antiquark structure of the pion was tested at Fermilab by Ashery et al.~\cite{Ashery:1999nq} by observing 500 GeV  pions dissociate into dijets $\pi + A \to q + \bar{q} + A' + X$ on a nuclear target.  Their work was motivated by theoretical studies of diffractive excitation in quantum chromodynamics \cite{Bertsch:1981py}, where diffraction arises from nuclear transparency to the projectile wave function with small transverse separation between the projectile's constituents~\cite{Aitala:2000hc}.

In this paper we propose an extension of the Ashery diffractive dissociation method to probe the fundamental structure of mesons, nucleons and nuclei.  We shall show how one can probe phenomena such as the quark-diquark structure of baryons, short-range nucleon-nucleon correlations and QCD ``hidden-color" \cite{Brodsky:1983vf} degrees of freedom inside nuclei.  The main goal is to
test the ``isophobic" nature of short-range correlations (SRCs) within nuclei analyzed by the CLAS collaboration at Jefferson Laboratory \cite{2018Natur.560..617C} and first reported in 2006 \cite{Piasetzky:2006ai}.  
The term isophobic refers to strong isospin quantum numbers, applies to both nucleon and quark doublets under $SU(2)_{\rm I}$ and is defined by $\Sigma ~\rm I=\Sigma ~\rm I_z=0$.  This means that for nucleon-nucleon interactions, proton-neutron interactions are highly favored over neutron-neutron or proton-proton; proton-neutron isospin quantum numbers sum to zero, thus the interaction is labeled ``isophobic."
Isophobic SRCs could provide a solution to the long-standing problem observed in deep inelastic scattering (DIS) experiments of leptons on nuclei, which show nuclear structure that does not follow simple additive cross section behavior \cite{Aubert:1983xm}, a phenomenon known as the ``EMC effect" \cite{Norton:2003cb}. 

We also discuss how diffractive dissociation of nuclei can test a new model for the isophobic nature of the EMC effect, in which short-range quark-quark correlations in nuclei are dominated by color-singlet 12-quark configurations: the ``hexa-diquark."  Formed out of nearest-neighbor scalar $[ud]$ diquarks, this multi-diquark model may also form strongly bound tetra-diquark $+$ valence quark states and thus may be visible in the MARATHON experiment on $A=3$ nuclei at JLab \cite{Petratos:2000qp}.

The EMC effect was first discovered in 1983 in DIS experiments over a wide range of nuclear targets as a strong distortion of the nuclear quark distributions in the domain of high momentum fraction:  Bjorken scaling parameter values of $0.3 < x_{B} < 0.7$.  Recently  the CLAS collaboration at Jefferson Laboratory~\cite{2018Natur.560..617C} compared nuclear parton distributions for a large range of nuclear targets and found, surprisingly, that the EMC effect has  a strongly isophobic nature. One model for this behavior is to identify short-range correlations between neighboring $n$ and $p$ nucleons which are much stronger compared to $p-p$ or $n-n$ SRCs.

In fact there are two conventional models to explain the EMC effect, either a modification of the mean field inside nuclei or the short-ranged 
correlated nucleon pairing within the nucleus~\cite{Weinstein:2010rt,Cloet:2019mql}.  The CLAS results are in favor of the nucleon SRC model and the data show the highly isophobic nature previously mentioned; similar nucleons are much less likely to pair than proton-neutron pairs at a factor of 20 times higher than proton-proton and neutron-neutron pairs \cite{2018Natur.560..617C,Subedi:2008zz}.  A definitive experimental test of this model is highly desirable and the diffractive dissociation of alpha particles is a straightforward and robust way to measure it.  

In an accompanying paper~\cite{West:2020rw} we suggest an alternative explanation for the EMC effect.  We postulate that the isophobic nature is due to high-virtuality quark and gluon dynamics underlying nuclear structure in terms of QCD short-distance degrees of freedom.  The formation of a $|[ud][ud][ud][ud][ud][ud]\rangle$ hexa-diquark, a charge-2, spin-0, baryon number-4, isospin-0, color-singlet intermediate state is proposed as an essential component of nuclear wavefunctions. Each $[ud]$ in the hexa-diquark is a scalar (spin-isospin singlet) diquark with $\bar 3_C$ color.  The hexa-diquark is a strongly-bound color-singlet configuration, and it is thus a natural component of the $\rm ^4 He$ nucleus wave function in terms of QCD degrees of freedom.  It may be related to color superconductivity models, e.g., of Alford  {\it et al.}~\cite{Alford:1997zt}, as it could form from an analog of a localized color-charged Cooper-pair condensate in the core of the nucleus.  A multi-diquark core, whether a strongly bound hidden-color state or a bosonic color-condensate, will allow heavier nuclei to form central multi-diquark structures as well.  We discuss implications for $A=3$ and other nuclei therein~\cite{West:2020rw}.

In this work, we show that the diffractive dissociation of relativistic nuclei can provide a definitive test of the isophobic nature of short-range correlations in the nuclear wave function as well as a way to discriminate between nucleon-based and quark-based models for the EMC effect.  Predictions made by light-front holographic QCD \cite{Brodsky:2014yha} may also be tested.

 Light-front wave functions (LFWFs)~\cite{Brodsky:1997de} are the eigenfunctions of the light-front Hamiltonian of relativistic quantum field theories quantized on the light front~\cite{Dirac:1949okn}.  They encode  the  properties of bound state systems such as  hadrons or nuclei in QCD and atoms in QED.  LFWFs are frame independent and causal and they underlie a wide range of bound-state observables: form factors, structure functions, distribution amplitudes, generalized parton distributions, transverse momentum dependent distributions, weak decays, final state interactions, and more.  Light-front holography provides a connection between an effective gravity theory in anti-de Sitter (AdS) space in five dimensions~\cite{Polchinski:2002jw} and a semiclassical approximation to color-confining QCD in $3+1$ spacetime dimensions~\cite{Brodsky:2006uqa, deTeramond:2008ht}.

Ashery developed a novel method to obtain the light-front wavefunction of the pion by measuring the diffractive dissociation of a relativistic pion into two jets upon a nuclear target
\bea
\pi+  A \to q + \bar{q} +  A',
\eea
where the high-momentum pion interacts with the nucleus with minimal momentum transfer~\cite{Ashery:1999nq}. When the target is left intact, the two-gluon exchange contribution between the high-momentum projectile and the target nucleus retains the color of the projectile (a color singlet for the pion) and gives an imaginary amplitude and an approximately constant diffractive cross section; this is characteristic of Pomeron exchange~\cite{Low:1975sv,Nussinov:1975mw}. To first approximation, the amplitude measures the second derivative of the pion light-front wave function \cite{Brodsky:1997de,Frankfurt:1999tq,Nikolaev:2000sh}
\bea
\frac{d \sigma}{d k_\perp^2} \propto  \left\vert \frac{\partial^2}{\partial k_\perp^2} \psi(x, \vec k_\perp) \right\vert.
\eea

The transverse impact parameter separation of the two quarks in the pion is inversely proportional to the relative transverse momenta.  After the soft interaction with the nuclei, two diffractive jets with opposite transverse momenta $\vec k_\perp$  and -$\vec k_\perp$ are produced, and their longitudinal momenta can be translated into the light front variable $x$, with one jet having $x$ and the other $(1-x)$. Therefore both the transverse and the longitudinal $x$-distributions of the pion can be obtained.  In the high $k_\perp^2$ regime, the hard gluonic interaction between the quark and the antiquark will be observed as diffractive jets.  Due to unitarity,  final state interactions can only produce an amplitude phase and therefore do not affect the proposed measurements.

A key feature of the diffractive dissociation process is ``color transparency'': When the pion dissociates to a quark and antiquark at high relative transverse momenta, $k^2_\perp \gg \Lambda^2_{\rm QCD}$,  the interaction on the nucleus is dominated by small-size color-singlet configurations \cite{Brodsky:1988xz}.  The pion is not absorbed and in fact interacts with each nucleon coherently \cite{Frankfurt:1993it}.

This mechanism can also be applied to other systems in order to obtain their constituent states. Let us consider other examples:

(1) {\it Diffractive dissociation of a relativistic proton}.  If the dominant Fock state configuration of the proton is three quarks $|uud\rangle$, one will most often observe 3 jets:  
\bea
p + A \to u~ +~ u +~ d + A'.
\eea
However, if the proton is dominated by quark-diquark configurations $|u [ud]\rangle$, as predicted by light-front holography and other models, one will instead observe two jets $p + A \to u + [ud] + A'$, similar to pion dissociation.  A subset of these events will also have $3$ jets when the dijet itself dissociates.  Light-front holography also predicts that the quark-diquark relative orbital angular momentum in the proton has equal probability for L=0 and L=1.  This could be tested by using a polarized proton beam and measuring the angular correlation of the quark and diquark jets.

(2) {\it Diffractive dissociation of relativistic nuclei to two nucleons.}  If the isophobic (I=0 np) nucleon-nucleon SRCs underlie the EMC effect, one would observe the diffractive dissociation of $\rm ^4He$:
\bea \label{npdiff}
\rm ^4He + A \to n + p + A' + X,
\eea
where the $n$ and $p$ have high relative transverse momenta.  In contrast, the I=1 nucleon pairs produced in
\bea   \label{ppdiff}
\rm ^4He + A \to p + p + A' + X
\eea
and 
\bea  \label{nndiff}
\rm ^4He + A \to n + n + A' + X 
\eea
events will have much smaller relative transverse momenta.  It is also possible to see the diffractive dissociation of $^4\rm{He}$ to two deuterons.

(3) {\it Diffractive dissociation of relativistic nuclei to hidden color states.}  On the other hand, if the $\alpha$  particle is dominated by the hexa-diquark configuration at short distances, diffractive dissociation of the $\alpha$ particle into multiple color jets is the primary signature.  An example of one such signature is three double-diquark jets, such as
\bea \label{jjdiff}
\rm ^4He+A\to\big[[ud][ud]\big]+\big[[ud][ud]\big]+ \big[[ud][ud]\big]+A'.~~~
\eea
These three diffractive jets would be J=0 double-diquarks each with color $3_C$.  Additional signatures of the hexa-diquark include multiple color jets with various combinations of diquarks, e.g., $2$ jets with $[ud][ud]$ and $[ud][ud][ud][ud]$ constituents.  This would be a $\bar{3} \times \bar{3} \rightarrow {3}$ color jet with a $(\bar{3}\times \bar{3}) \times (\bar{3} \times \bar{3}) \rightarrow 3 \times 3 \rightarrow \bar{3}$ color jet.  At very small rates, the helium nucleus would dissociate into all $6$ diquarks  with a $6$ color jet signature.  These are all quite distinct from $n-p$ correlation signatures.  Single quark jets would be suppressed compared to diquark jets in this model since all quarks are bound into diquarks.

(4) {\it Diffractive dissociation of relativistic deuterium nuclei to hidden-color systems.} The deuteron is a J=1, I=0 color singlet bound state in QCD and will normally dissociate into $n-p$ states due to its weak binding energy.  However, at small transverse separation between constituents the deuteron is about 80\% hidden color from different color-singlet combinations of its six quarks~\cite{Brodsky:1983vf}. Diffractive dissociation measurements can help disentangle the short-distance behavior of the deuteron wave function: For example, 4 diffractive color jets $3 + 3 + \bar3 + \bar 3$ in
\bea
d + A \to u + d + [ud] + [ud] + A'.
\eea
The deuteron is special in the hexa-diquark model as it can make use of the $6\times\bar{6}$ singlet combination of the two quark-diquark structures in the nucleus.  Since diquarks are in the $\bar{3}$ of color, quarks in the $3$,  diquark-diquark bindings between the proton and neutron give $\bar{3}\times\bar{3}\to 3+\bar{6}$.  The valence quarks combine as $3\times3\to \bar{3}+6$, thus opening up a $6\times\bar{6}\to 1$ combination, a fact that may explain the deuteron's weak binding energy of $2.2~ \rm MeV$  \cite{West:2020rw}.

We note that a diffracted color-singlet hexaquark (tri-diquark) such as $[ud][ud](ud)$, where the spin $S=1$ diquark $(ud)$ is a member of the isospin $I = 1$ triplet $(uu), (ud), (dd)$ appears to be excluded by spin-statistics. Such state would not be a natural Fock component of the deuteron wave function, which is spin triplet but isospin singlet.  The deuteron Fock state can also be matched to an isobar pair such as $\Delta^{++} (uuu)$ with $\Delta^-(ddd)$.  The weight of each configuration can be measured by diffractive dissociation, such as 
\bea
d + A \to \Delta^{++} + \Delta^- + A'.
\eea

5) {\it Diffractive dissociation of relativistic electromagnetic systems.}  One could measure the light-front wavefunction of positronium and other atoms through diffractive dissociation of in-flight atoms. In this case, the Coulomb one-photon exchange measures the first derivative of the positronium light-front wavefunction.  In the case of pion diffractive dissociation to dijets with large transverse momentum $k_\perp$,  the  two-gluon exchange amplitude is obtained by replacing $b^2_\perp$ in the dipole amplitude by the second derivative of the light-front wave function with respect to $k_\perp$ 
\cite{Ashery:2006zw}, where $b_\perp$ is the 
transverse size of the $\bar{q}q$ system. 
In the case of the coherent photoproduction of di-muons, 
the dominant amplitude is given by one-photon exchange, 
rather than two-gluons, and is therefore obtained from the first derivative 
of the light-front wavefunction with respect to $k_\perp$, not the second derivative.

It is possible to have minimal momentum transfer to the nuclear target, which is left intact. This would separate and measure the two most important Fock states in positronium, $(e^+ e^-)$ and $(e^+ e^-  \gamma)$.
A nonperturbative analysis of these light-front wave functions was performed in 2014~\cite{Wiecki:2014ola}.

The diffractive cross section for Coulomb dissociation of the atom on the nucleus is proportional to $\rm Z^2$. Thus the target need not be ionized atoms, any large enough Z nuclear target, e.g., even carbon, will do.  The electron and positron in positronium will each interact coherently, via photon exchange, with the nucleus and the cross section is dominated by the bound Z-proton system by the factor of $\rm Z^2$.  When the electron and positron interact with valence electrons, the cross section is suppressed with respect to the nucleus by the same factor of $\rm Z^2$ because they interact with a single valence electron at a time.  

Coulomb dissociation of true muonium $(\mu^+ \mu^-)$ would be particularly interesting \cite{Brodsky:2009gx}.  The transition of the momentum dependence of its light-front wave-function~\cite{Lamm:2016djr} from 
\bea
\frac{1}{k_\perp^4} ~\rm \to~\frac{1}{k_\perp^2},
\eea
from the Coulomb dominated non-relativistic behavior ($1/k_\perp^4$) to the relativistic domain with a slower fall-off ($1/k_\perp^2$), is a primary experimental signature~\cite{Greub:1994fp,Brodsky:2011fc}.

Several muon related experimental results may point to physics beyond the Standard Model.  These include the anomalous magnetic moment of the muon~\cite{Bennett:2006fi}, lepton non-universality in B meson decays~\cite{Ciezarek:2017yzh,Aaij:2019bzx} and possibly Higgs decays, although early hints in the Higgs sector~\cite{Khachatryan:2015kon} appear to have gone away~\cite{Sirunyan:2017xzt}.  The analysis and detection of true muonium Fock states may be of great interest for exploring solutions to these puzzles \cite{Lamm:2015fia} as it tests whether the muon truly behaves as a more massive version of the electron.

In this article, we have proposed a set of diffractive dissociation experiments to investigate the fundamental structure of hadrons and nuclei. In particular, the diffractive dissociation of relativistic $\alpha$-particles to two nucleons on a nuclear target $A$ constitutes a definitive test of the isospin dependence of strongly correlated nucleon pairs in nuclei: Diffractive $n-p$ pairs in (\ref{npdiff}) would have high relative transverse momenta compared with non-diffractive $p-p$ or $n-n$ pairs in (\ref{ppdiff}) or (\ref{nndiff}). On the other hand, the observation of two or more color diffractive jets as in (\ref{jjdiff}), would test short-range QCD dynamics in terms of quarks and gluons, such as hidden-color diquarks in the nuclear wave function.

We conclude with specific 
recommendations for diffractive experiments aimed at disentangling the nature of the SRC nucleon pairs in nuclei.
There are two steps to any such experiment:

1.  Verify that the experiment is probing the Feynman scaling regime, i.e., that the differential cross section for the leading particles is energy independent.  This requires at least $2$ measurements of the cross section at different energies and in fact $2$ would be sufficient.  The cross section should be independent of the beam energy.  In this regime we are probing the constituents of the beam particles.  As a specific example, with a $4~ \rm GeV$ $\rm ^4He$ beam upon a nuclear target, measurement of the cross section at $4~ \rm GeV$ and $3.5 ~\rm GeV$ yielding the same results verifies Feynman scaling \cite{Feynman:1969ej}.

2.  Measure the relative transverse momenta of the dissociated products, that is 
\bea 
\Delta_{k_\perp}^{\rm AB}=k_{\perp}^{\rm A}-k_{\perp}^{\rm B},
\eea
and identify the particles $\rm A$ and $\rm B$ for which $\Delta_{k_\perp}^{\rm AB}$ is maximized.  For the isophobic SRC pairs in the EMC effect experiments, $\rm A=proton$, $\rm B=neutron$ will yield the maximal value by a factor of $20$ to $1$ if the CLAS results are to be confirmed.


\subsection*{Acknowledgements}
We thank J. D. Bjorken and Or Hen for helpful discussions.  This work is supported in part by the Department of Energy, Contract DE--AC02--76SF00515. IS is supported by Fondecyt (Chile) grant No. 1180232 and CONICYT PIA/BASAL FB0821. SLAC-PUB-17500.



\begin{thebibliography}{50}

\bibitem{Ashery:1999nq} 
  D.~Ashery {\it et al.} [E791 Collaboration],
  ``Diffractive dissociation of high momentum pions,"
  Bloomington 1999, Physics with a high luminosity polarized electron ion collider, p. 322-334
  [\href{https://arxiv.org/abs/hep-ex/9910024}{\tt hep-ex/9910024}];
  
  E.~M.~Aitala {\it et al.} [E791 Collaboration],
  ``Direct measurement of the pion valence-quark momentum distribution, the pion light-cone wave function squared,''
  \href{https://journals.aps.org/prl/abstract/10.1103/PhysRevLett.86.4768} {Phys.\ Rev.\ Lett.\ {\bf 86}, 4768 (2001)}
  [\href{https://arxiv.org/abs/hep-ex/0010043}{\tt hep-ex/0010043}].
  
  
\bibitem{Bertsch:1981py} 
  G.~Bertsch, S.~J.~Brodsky, A.~S.~Goldhaber and J.~F.~Gunion,
  ``Diffractive excitation in QCD,"
  \href{https://journals.aps.org/prl/abstract/10.1103/PhysRevLett.47.297}{Phys.\ Rev.\ Lett.\  {\bf 47}, 297 (1981)}.
  

\bibitem{Aitala:2000hc} 
  E.~M.~Aitala {\it et al.} [E791 Collaboration],
  ``Observation of color transparency in diffractive dissociation of pions,''
  \href{https://journals.aps.org/prl/abstract/10.1103/PhysRevLett.86.4773}{Phys.\ Rev.\ Lett.\  {\bf 86}, 4773 (2001)}
  [\href{https://arxiv.org/abs/hep-ex/0010044}{\tt hep-ex/0010044}].
  

\bibitem{Brodsky:1983vf} 
  S.~J.~Brodsky, C.~R.~Ji and G.~P.~Lepage,
  ``Quantum chromodynamic predictions for the deuteron form factor,"
  \href{https://journals.aps.org/prl/abstract/10.1103/PhysRevLett.51.83}{Phys.\ Rev.\ Lett.\  {\bf 51}, 83 (1983)}.


  
\bibitem{2018Natur.560..617C}
       CLAS Collaboration,
       ``Probing high-momentum protons and neutrons in neutron-rich nuclei,"
       \href{https://www.nature.com/articles/s41586-018-0400-z}{Nature {\bf 560}, 617 (2018)}.
       
\bibitem{Piasetzky:2006ai} 
  E.~Piasetzky, M.~Sargsian, L.~Frankfurt, M.~Strikman and J.~W.~Watson,
  ``Evidence for the strong dominance of proton-neutron correlations in nuclei,''
  \href{https://journals.aps.org/prl/abstract/10.1103/PhysRevLett.97.162504}{Phys.\ Rev.\ Lett.\  {\bf 97}, 162504 (2006)}
  [\href{https://arxiv.org/abs/nucl-th/0604012}{nucl-th/0604012]}.
      
\bibitem{Aubert:1983xm} 
  J.~J.~Aubert {\it et al.} [European Muon Collaboration],
  ``The ratio of the nucleon structure functions $F_2^N$ for iron and deuterium,"
  \href{https://www.sciencedirect.com/science/article/abs/pii/0370269383904379?via%3Dihub}{Phys.\ Lett.\ {\bf 123B}, 275 (1983).}
 
  
   \bibitem{Norton:2003cb} 
  P.~R.~Norton,
  ``The EMC effect,"
  \href{https://iopscience.iop.org/article/10.1088/0034-4885/66/8/201}{Rept.\ Prog.\ Phys.\  {\bf 66}, 1253 (2003)}.  
  
\bibitem{Petratos:2000qp} 
  G.~G.~Petratos, I.~R.~Afnan, F.~R.~P.~Bissey, J.~Gomez, A.~T.~Katramatou, W.~Melnitchouk and A.~W.~Thomas,
  ``Measurement of the $F_2^n/F_2^p$ and $d/u$ ratios in deep inelastic electron scattering off $\rm^3H$ and $\rm^3He$,"
  \href{https://arxiv.org/abs/nucl-ex/0010011}{\tt nucl-ex/0010011}.

  
 \bibitem{Weinstein:2010rt} 
  L.~B.~Weinstein, E.~Piasetzky, D.~W.~Higinbotham, J.~Gomez, O.~Hen and R.~Shneor,
  ``Short range correlations and the EMC effect,"
  \href{https://journals.aps.org/prl/abstract/10.1103/PhysRevLett.106.052301}{Phys.\ Rev.\ Lett.\  {\bf 106}, 052301 (2011)}
  [\href{https://arxiv.org/abs/1009.5666}{\tt arXiv:1009.5666 [hep-ph]}]. 
  
  
\bibitem{Cloet:2019mql} 
  I.~C.~Clo\"et {\it et al.},
   ``Exposing novel quark and gluon effects in nuclei,"
 \href{https://iopscience.iop.org/article/10.1088/1361-6471/ab2731}{J. Phys.G {\bf 46}, no. 9, 093001 (2019)}
  \href{https://arxiv.org/abs/1902.10572}{[\tt arXiv:1902.10572 [nucl-ex]}].
  
  
\bibitem{Subedi:2008zz} 
  R.~Subedi {\it et al.},
  ``Probing cold dense nuclear matter,"
  \href{http://science.sciencemag.org/content/320/5882/1476}{Science {\bf 320}, 1476 (2008)}
  [\href{https://arxiv.org/abs/0908.1514}{\tt arXiv:0908.1514 [nucl-ex]}].
  
  
\bibitem{West:2020rw}
 J.~Rittenhouse~West, S.~J.~Brodsky, G.~F.~de~T\'eramond, A.~Goldhaber, I.~Schmidt,
 ``QCD hidden-color diquark dynamics in the central core of nuclei,"
 {\tt arXiv:2001.XXXX [hep-ph]]}.
 
 
\bibitem{Alford:1997zt} 
  M.~G.~Alford, K.~Rajagopal and F.~Wilczek,
  ``QCD at finite baryon density: Nucleon droplets and color superconductivity,"
 \href{https://www.sciencedirect.com/science/article/abs/pii/S0370269398000513?via%3Dihub}{Phys.\ Lett.\ B {\bf 422}, 247 (1998)}
  [\href{https://arxiv.org/abs/hep-ph/9711395}{\tt hep-ph/9711395}].


\bibitem{Brodsky:2014yha} 
  S.~J.~Brodsky, G.~F.~de T\'eramond, H.~G.~Dosch and J.~Erlich,
  ``Light-front holographic QCD and emerging confinement,"
  \href{https://www.sciencedirect.com/science/article/abs/pii/S0370157315002306?via%3Dihub}{Phys.\ Rept.\  {\bf 584}, 1 (2015)}
  [\href{http://arxiv.org/abs/arXiv:1407.8131}{\tt arXiv:1407.8131 [hep-ph]}].


\bibitem{Brodsky:1997de} 
  S.~J.~Brodsky, H.~C.~Pauli and S.~S.~Pinsky,
  ``Quantum chromodynamics and other field theories on the light cone,"
  \href{https://www.sciencedirect.com/science/article/abs/pii/S0370157397000896?via%3Dihub}{Phys.\ Rept.\  {\bf 301}, 299 (1998)}
  [\href{https://arxiv.org/abs/hep-ph/9705477}{\tt hep-ph/9705477}].
  
  
\bibitem{Dirac:1949okn} 
 P.~A.~M.~Dirac,
 ``Forms of relativistic dynamics,"
 \href{https://journals.aps.org/rmp/abstract/10.1103/RevModPhys.21.392}{Rev.\ Mod.\ Phys.\  {\bf 21}, 392 (1949)}.
 
 
 
\bibitem{Polchinski:2002jw} 
  J.~Polchinski and M.~J.~Strassler,
  ``Deep inelastic scattering and gauge/string duality,"
  \href{https://journals.aps.org/prl/abstract/10.1103/PhysRevLett.88.031601}{JHEP {\bf 0305}, 012 (2003)}
  [{\href{http://arxiv.org/abs/hep-th/0109174}{\tt hep-th/0209211}].
  

\bibitem{Brodsky:2006uqa} 
  S.~J.~Brodsky and G.~F.~de T\'eramond,
 ``Hadronic spectra and light-front wavefunctions in holographic QCD,''
  \href{https://journals.aps.org/prl/abstract/10.1103/PhysRevLett.96.201601}{Phys.\ Rev.\ Lett.\  {\bf 96}, 201601 (2006)}
  [\href{http://arxiv.org/abs/hep-ph/0602252}{\tt hep-ph/0602252}].
  

 \bibitem{deTeramond:2008ht} 
  G.~F.~de T\'eramond and S.~J.~Brodsky,
  ``Light-front holography: A first approximation to QCD,''
  \href{https://journals.aps.org/prl/abstract/10.1103/PhysRevLett.102.081601}{Phys.\ Rev.\ Lett.\  {\bf 102}, 081601 (2009)}
  [\href{http://arxiv.org/abs/arXiv:0809.4899}{\tt arXiv:0809.4899 [hep-ph]}].

\bibitem{Low:1975sv} 
  F.~E.~Low,
  ``A model of the bare Pomeron,''
  \href{https://journals.aps.org/prd/abstract/10.1103/PhysRevD.12.163}{Phys.\ Rev.\ D {\bf 12}, 163 (1975)}.
  
\bibitem{Nussinov:1975mw} 
  S.~Nussinov,
  ``Colored-quark version of some hadronic puzzles,''
  \href{https://journals.aps.org/prl/abstract/10.1103/PhysRevLett.34.1286}{Phys.\ Rev.\ Lett.\  {\bf 34}, 1286 (1975)}.
  }


\bibitem{Frankfurt:1999tq} 
  L.~Frankfurt, G.~A.~Miller and M.~Strikman,
  ``Perturbative pion wave function in coherent pion-nucleon dijet production,"
  \href{https://link.springer.com/article/10.1023%2FA%3A1003616828027}{Found.\ Phys.\  {\bf 30}, 533 (2000)}
  [\href{http://arxiv.org/abs/hep-ph/9907214}{\tt hep-ph/9907214}].

\bibitem{Nikolaev:2000sh} 
  N.~N.~Nikolaev, W.~Schafer and G.~Schwiete,
  ``Coherent production of hard dijets on nuclei in QCD,"
  \href{https://journals.aps.org/prd/abstract/10.1103/PhysRevD.63.014020}{Phys.\ Rev.\ D {\bf 63}, 014020 (2001)}
  [\href{https://arxiv.org/abs/hep-ph/0009038}{\tt hep-ph/0009038}].

\bibitem{Brodsky:1988xz} 
  S.~J.~Brodsky and A.~H.~Mueller,
  ``Using nuclei to probe hadronization in QCD,"
  \href{https://www.sciencedirect.com/science/article/abs/pii/0370269388907198?via%3Dihub}{Phys.\ Lett.\ B {\bf 206}, 685 (1988)}.

 
\bibitem{Frankfurt:1993it} 
  L.~Frankfurt, G.~A.~Miller and M.~Strikman,
  ``Coherent nuclear diffractive production of minijets -- illuminating color transparency,"
  \href{https://www.sciencedirect.com/science/article/abs/pii/0370269393913909?via%3Dihub}{Phys.\ Lett.\ B {\bf 304}, 1 (1993)}
  [\href{http://arxiv.org/abs/hep-ph/9305228}{\tt hep-ph/9305228}].

\bibitem{Ashery:2006zw}
D.~Ashery,
``High momentum diffractive processes and hadronic structure,''
\href{https://www.sciencedirect.com/science/article/abs/pii/S0146641005000992?via%3Dihub}{Prog.\ Part.\ Nucl.\ Phys.\  \textbf{56}, 279-339 (2006)}


\bibitem{Wiecki:2014ola} 
  P.~Wiecki, Y.~Li, X.~Zhao, P.~Maris and J.~P.~Vary,
  ``Basis light-front quantization approach to positronium,"
  \href{https://journals.aps.org/prd/abstract/10.1103/PhysRevD.91.105009}{Phys.\ Rev.\ D {\bf 91},  105009 (2015)}
  [\href{https://arxiv.org/abs/1404.6234}{\tt arXiv:1404.6234 [nucl-th]}].

\bibitem{Brodsky:2009gx} 
  S.~J.~Brodsky and R.~F.~Lebed,
  ``Production of the smallest QED atom: True muonium ($\mu^+ \mu^-$),"
  \href{https://journals.aps.org/prl/abstract/10.1103/PhysRevLett.102.213401}{Phys.\ Rev.\ Lett.\  {\bf 102}, 213401 (2009)}
  [\href{https://arxiv.org/abs/0904.2225}{\tt arXiv:0904.2225 [hep-ph]}].

  
\bibitem{Lamm:2016djr} 
  H.~Lamm and R.~Lebed,
  ``High resolution nonperturbative light-front simulations of the true muonium atom,''
  \href{https://journals.aps.org/prd/abstract/10.1103/PhysRevD.94.016004}{Phys.\ Rev.\ D {\bf 94}, no. 1, 016004 (2016)}
  [\href{https://arxiv.org/abs/1606.06358}{\tt arXiv:1606.06358 [hep-ph]]}.



\bibitem{Greub:1994fp} 
  C.~Greub, D.~Wyler, S.~J.~Brodsky and C.~T.~Munger,
  ``Atomic alchemy: Weak decays of muonic and pionic atoms into other atoms''
  \href{https://journals.aps.org/prd/abstract/10.1103/PhysRevD.52.4028}{Phys.\ Rev.\ D {\bf 52}, 4028 (1995)}
  [\href{https://arxiv.org/abs/hep-ph/9405230}{\tt hep-ph/9405230}].

\bibitem{Brodsky:2011fc} 
  S.~J.~Brodsky,
  ``Atoms in flight and the remarkable connections between atomic and hadronic physics,''
  \href{https://link.springer.com/article/10.1007%2Fs10751-012-0576-9}{Hyperfine Interact.\  {\bf 209}, no. 1-3, 83 (2012)}
  [\href{https://arxiv.org/abs/1112.0628}{\tt arXiv:1112.0628 [hep-ph]}].

\bibitem{Bennett:2006fi} 
  G.~W.~Bennett {\it et al.} [Muon g-2 Collaboration],
  ``Final report of the muon E821 anomalous magnetic moment measurement at BNL,"
  \href{https://journals.aps.org/prd/abstract/10.1103/PhysRevD.73.072003}{Phys.\ Rev.\ D {\bf 73}, 072003 (2006)}
  [\href{https://arxiv.org/abs/hep-ex/0602035}{\tt hep-ex/0602035}].
  
\bibitem{Ciezarek:2017yzh} 
  G.~Ciezarek, M.~Franco Sevilla, B.~Hamilton, R.~Kowalewski, T.~Kuhr, V.~Lüth and Y.~Sato,
  ``A challenge to lepton universality in B meson decays,"
  \href{https://www.nature.com/articles/nature22346}{Nature {\bf 546}, 227 (2017)}
  [\href{https://arxiv.org/abs/1703.01766}{\tt arXiv:1703.01766 [hep-ex]}].

\bibitem{Aaij:2019bzx} 
  R.~Aaij {\it et al.} [LHCb Collaboration],
  ``Test of lepton universality with $\Lambda^{0}_{b} \to p K^- \ell^+ \ell^-$ decays,"
  [\href{https://arxiv.org/abs/1912.08139}{\tt arXiv:1912.08139 [hep-ex]}].

\bibitem{Khachatryan:2015kon} 
  V.~Khachatryan {\it et al.} [CMS Collaboration],
  ``Search for lepton-flavour-violating decays of the higgs boson,"
  \href{https://www.sciencedirect.com/science/article/pii/S0370269315005638?via%3Dihub}{Phys.\ Lett.\ B {\bf 749}, 337 (2015)}
  [\href{https://arxiv.org/abs/1502.07400}{\tt arXiv:1502.07400 [hep-ex]}].

\bibitem{Sirunyan:2017xzt} 
  A.~M.~Sirunyan {\it et al.} [CMS Collaboration],
  ``Search for lepton flavour violating decays of the Higgs boson to $\mu\tau$ and e$\tau$ in proton-proton collisions at $\sqrt{s}=$ 13 TeV,"
  \href{https://link.springer.com/article/10.1007%2FJHEP06%282018%29001}{JHEP {\bf 1806}, 001 (2018)}
  [\href{https://arxiv.org/abs/1712.07173}{\tt arXiv:1712.07173 [hep-ex]}].
  



\bibitem{Lamm:2015fia} 
  H.~Lamm,
  ``True muonium: The atom that has it all,''
  [\href{https://arxiv.org/abs/1509.09306}{\tt arXiv:1509.09306 [hep-ph]}].

\bibitem{Feynman:1969ej} 
  R.~P.~Feynman,
  ``Very high-energy collisions of hadrons,''
  \href{https://journals.aps.org/prl/abstract/10.1103/PhysRevLett.23.1415}{Phys.\ Rev.\ Lett.\  {\bf 23}, 1415 (1969)}.
  

  

\end{thebibliography}
\end{document}